\documentclass{ws-ijmpd}
\def\be{\begin{equation}}
\def\ee{\end{equation}}
\def\bc{\begin{center}}
\def\ec{\end{center}}

\newcommand{\bea}{\begin{eqnarray}}
\newcommand{\eea}{\end{eqnarray}}
\begin{document}

\markboth{A.F. Zakharov,  V.N. Pervushin}
{Conformal Cosmological Model Parameters with Distant SNe Ia Data}

%
\catchline{}{}{}{}{}
%

\title{CONFORMAL COSMOLOGICAL MODEL PARAMETERS WITH DISTANT SNe Ia DATA: "GOLD" AND "SILVER"
}

\author{ALEXANDER F. ZAKHAROV}
\address{National Astronomical Observatories of Chinese Academy of
Sciences, \\20A Datun Road, Chaoyang District, Beijing, 100012, China;\\
Institute of Theoretical and Experimental Physics, \\
B. Cheremushkinskaya, 25,
117259, Moscow, Russia; \\
Bogoliubov Laboratory for Theoretical Physics, \\
JINR, 141980 Dubna,
Russia\\
zakharov@itep.ru}
%


\author{VICTOR N. PERVUSHIN}
\address{Bogoliubov Laboratory for Theoretical Physics, \\
JINR, 141980 Dubna,
Russia\\
pervush@theor.jinr.ru}

\maketitle

\begin{history}
\received{Day Month Year}
\revised{Day Month Year}
\end{history}

\begin{abstract}
Assuming that supernovae type Ia (SNe Ia) are standard candles one
could use them to test cosmological theories. The Hubble Space
Telescope team analyzed  186 SNe Ia\cite{Riess_04} to test the
Standard Cosmological model (SC) associated with expanded lengths
in the Universe and evaluate its parameters. We use the same
sample to determine parameters of Conformal Cosmological model
(CC) with relative reference units of intervals, so that conformal
quantities of General Relativity are interpreted as observables. We concluded, that really the
test is extremely useful and allows to evaluate parameters of the
model. From a formal statistical point of view the best fit of the
CC model is almost the same quality approximation as  the best fit
of SC model with $\Omega_\Lambda=0.72, \Omega_m=0.28$. As it was
noted earlier, for CC models, a rigid matter component could
substitute the $\Lambda$-term (or quintessence) existing in the SC
model. We note that a free massless scalar field can generate
such a rigid matter. We describe  results of our analysis for more recent
"gold" data (for 192 SNe Ia).

\end{abstract}


\section{Introduction}

Now there is enormous progress in observational and theoretical
cosmology and even it is typically accepted that cosmology enters
into an era of precise science (it means that a typical accuracy of
standard parameter determination is about few percents), despite,
there are different approaches including alternative theories of
gravity to fit observational data (see recent reviews\cite{Will_2006}
for references). Some classes of theories could be
constrained by Solar system data\cite{Zakharov_06} even if they
passed cosmological tests. Thus, all the theories should pass all
possible tests including cosmological ones.

Since the end of the last century distant supernovae data is a
widespread test for all theoretical cosmological models in spite
of the fact the correctness of the hypothesis about SNe Ia as the
perfect standard candles is still not proven.\cite{Panagia_05}
However, the first observational conclusion about accelerating
Universe and existence of non-vanishing $\Lambda$-term was done
with the cosmological SNe Ia data. Therefore, typically  standard
(and alternative) cosmological approaches are checked with the
test.

Conformal cosmological models \cite{Behnke_02,Behnke_04,CC_papers},
where all observables are identified with the scale-invariant quantities
of GR introduced
 yet by Lichnerowicz \cite{lich},
are also discussed among other possibilities.\cite{Riess_2001}
 {~The
Conformal Cosmology
is based on the Dirac version of the GR.\cite{Dirac_73} Dirac  modified the  accepted General Relativity (GR) in
spirit of the simplified Weyl's geometry,\cite{weyl}
which means that
``\textit{a new action principle was set up, much simpler than Weyl's, but
requiring
  a scalar field function}'' (called here as dilaton)
``\textit{to describe the gravitation field, in additional to
$g_{\mu\nu}$}''\cite{Dirac_73}.
  The Dirac version of GR
\be \label{dm-1}S_{\rm Dirac}=-\phi_0^2\int d^4x
\left[\dfrac{e^{-2D}\sqrt{-\widetilde{g}}}{6}R^{(4)}(\widetilde{g})+
e^{-D}\partial_\mu\left(\sqrt{-\widetilde{g}}\widetilde{g}^{\mu\nu}
\partial_\nu  e^{-D}\right)\right]\ee
is compatible with the choice of the Lichnerowicz variables
$|\widetilde{g}^{(3)}|=1$\cite{lich} as measurable ones that scale all
masses. The action
(\ref{dm-1}) is scale-invariant in contrast to the Brans -- Dicke theory.\cite{Faraoni_99}

The
Conformal Cosmology   is based on
   the Weyl definition\cite{weyl} of the measurable interval
 as the ratio of the Einstein interval and units defined as reversed masses
\be\label{sc-cc}
1+z=\frac{\lambda_0~~~m_0}{\left[\lambda_0
a(t)\right]m_0}=\frac{\lambda_0~~~m_0}{\lambda_0
\left[a(t)m_0\right]}, \ee where $\lambda_0$ is the wave length of
a photon emitted at the present-day instant and $m_0$ is the
standard mass defining the units of measurements. This Weyl
definition of the measurable interval gives a possibility to
consider two alternatives: the Standard Cosmology (SC)
 \be\label{S-1}(1+z)_{\rm sc}=\dfrac{\lambda_0}{\left[\lambda_0
a(t)\right]}\ee if  $a$ is jointed  to a length $\lambda_0$ (that
means expanded lengths in a universe), or the Conformal
Cosmology  (CC)
\be\label{C-1}(1+z)_{\rm cc}=\dfrac{m_0}{\left[a(t)m_0\right]}\ee
if we joint $a$ to a
 mass $m_0$ (that means running masses). The construction of all
 observable CC-quantities is based on the {\it conformal
 postulate} in accord to which all observable CC-quantities $F_c^{(n)}$
 with conformal weight
 $(n)$ are
 equal to the SC ones $F_s^{(n)}$ multiplied by the cosmological
 scale factor to the power $(-n)$
 \be\label{c-s}
 F_c^{(n)}=a^{-n}F_s^{(n)}
 \ee
  In accord with the conformal
 postulate (\ref{c-s}) the CC-time is greater than the SC one, and all
 CC-distances, including the CC-luminosity distance $\ell_c$, are longer that
 than the SC-ones $\ell_s=a\ell_c$, because all intervals are measured by
 clocks of mass {\it const}$/a$.

Conformal symmetry means that the really measurable quantity is the ratio
$[M_e  L_e]/[M_0  L_0]=a_e$
where $[M_e  L_e]$ is the conformal-invariant product of mass of the atom $M_e$
(reflecting its size),
 and  the wave-length $L_e$ of the atom photon, at the time of emission (e) ,
and
 $[M_0  L_0]$ is the similar product value at present day.

In the  papers \cite{Zee_79} devoted to the applications of the conformal
 symmetry in order to
 study and calculate  the short-distance effects in quantum gravity
 one counted that the source of  the cosmological scale factor ($a_e$) growth
  is the expansion of the lengths (i.e. $L_e  = a_e L_0, M_e
 = M_0$).
  In contrast to these papers,  in our approach  we select another possible
 alternative: (i.e. $M_e  = a_e M_0, L_e  =  L_0$).

 First attempts to analyze SNe Ia data to
evaluate parameters of CC models were done,\cite{Behnke_02} so it
was used only 42 high redshift type Ia SNe\cite{Riess_98} but
after that it was analyzed a slightly extended sample.\cite{Behnke_04}
 In spite of a small size of the samples used in previous attempts
 to fit CC model parameters, it was
concluded that if $\Omega_{\rm rig}$ is significant in respect to
the critical density, CC models could fit SNe Ia observational data
with a reasonable accuracy.
After that a possibility to fit observational SNe Ia data with CC models was
seriously discussed
by different authors among other alternatives.\cite{Riess_2001}

An aim of the paper is to check and
clarify previous conclusions about possible bands for CC parameters
with a more extended  (and more accurate) sample\cite{Riess_04}
used commonly to check standard and alternative cosmological models.
The HST cosmological SNe Ia team have corrected data of previous
smaller samples as well and also considered possible
non-cosmological but astronomical ways to fit observational ways and
concluded that some of them such a replenishing dust (with
$\Omega_m=1, \Omega_\Lambda=0.$) could fit observational data pretty
well even in respect to the best fit cosmological model.

The content of the paper
 is the following.
 In Section 2, the basic CC relations are reminded.
 In Section 3, a magnitude-redshift relation for distant SNe is
 discussed. In Section 4,
 results of fitting procedure for CC models with the "gold" and "silver" sample
 are given.  Conclusions are presented in Section 5.

\section{\label{s-1} Conformal Cosmology Relations}
We will remind basic relations for CC model parameters (see
papers\cite{Behnke_02,CC_papers} for details) considering the
General Relativity  with an additional  scalar field
$Q$, as usually people did to introduce quinessence\cite{Quintessence} (earlier, the approach was used for inflationary cosmology, see for example, paper\cite{Linde_07} and references therein)
\begin{eqnarray}\label{pct1}
S&=S_{\rm Dirac}+&\int\limits_{ }^{
}d^4x\sqrt{-g}\left[\frac{1}{2}\partial_\mu Q\partial^\mu Q-V(Q)\right];
\end{eqnarray}
here we used the natural units
 \be\label{nu-1}
 M_{\rm
 Planck}\sqrt{\frac{3}{8\pi}}=\hbar=c=1,
 \ee
therefore,
we have the following expressions for density and pressure of the scalar field ($p_Q$ and $\rho_Q$, respectively)\cite{Quintessence}
 \bea
 \label{density-1}
p_Q(t)=\frac{1}{2}\dot{Q}^2+V(Q),
\nonumber\\
\rho_Q(t)=\frac{1}{2}\dot{Q}^2-V(Q),
 \eea
and equation of state (EOS) such as $p_Q=w_Q\rho_Q$, where
\bea
\label{density-2}
w_Q=\dfrac{\frac{1}{2}\dot{Q}^2-V(Q)}{\frac{1}{2}\dot{Q}^2+V(Q)},
 \eea
($-1\le w_Q \le 1,$ for "natural" potentials $V(Q) \geq 0$).
In contrast with quintessence model where one uses typically $\dot{Q}^2\ll V(Q)$, below for CC model we will use an approximation
$\dot{Q}^2\gg V(Q)$ (for a standard representation of the potential $V(Q)=\frac{1}{2}m Q^2$, where $m$ is a mass of the field,
the approximation corresponds to
a massless field model) and we have
\be
\label{density-3}
w_Q=\frac{\frac{1}{2}\dot{Q}^2}{\frac{1}{2}\dot{Q}^2}=1,
 \ee
or on the other words, a rigid EOS for the scalar field  $p_Q=\rho_Q$ ($p_{\rm rig}=\rho_{\rm rig}$, since for our future needs an origin
of the EOS is not important, hereafter, we will call the component such as the rigid matter).


%
 The  conformal postulate means that CC-intervals
  \be\label{cc-1}
 ds^2_{\rm c}=\frac{ds^2_{\rm s}}{a^2}=(d\eta)^2-(dx^k)^2,
  \ee
  are greater than SC-intervals,
    CC-time $t_c=\eta=\int dt_s/a$ is greater than SC-time $t_s=t$
  CC-luminosity-distance
 \be\label{cc-1ell} \ell_c=\frac{\ell_s}{a},
 \ee
 is longer than the SC-one, conformal masses scaled by the factor $a$
 \be\label{cc-1m}
 m_{\rm c}=m_0a(\eta)
 \ee
 are less than constant SC masses $m_s=m_0$,
and a constant conformal temperature
 $T_c=aT_s=T_0$ is less that the SC-temperature $T_s=T_0/a$.

 In   homogeneous approximation both  SC and CC  are
described by
 \be\label{cce}
  (a')^2=\rho_c(a)
 \ee
 where $a'$ is the derivative of the cosmological scale factor
  $a$ with respect to  conformal time,
 \be\label{rh}
 \rho_c(a) =\rho_0
 \sum\limits_{J=-2,0,1,4}^{}\Omega_{J}a^J
 \ee is the conformal
energy density  connected with the SC one by the transformation
\be \rho_c(a)=a^4~\rho_{s}(a)~,\ee and $\Omega_{J}$ is partial
energy density
  marked by index $J$ running a set of values
   $J=-2$ (rigid), $J=0$ (radiation), $J=1$ (mass), and $J=4$ ($\Lambda$-term)
 in correspondence with a type of matter field contributions; here
 $\sum\limits_{J=-2,0,1,4}^{}\Omega_{J}=1$ is assumed.
 The case $J=-2$ corresponds to
 a rigid state, where the energy  density coincides with the
 pressure one
 $\rho=p$. The rigid state can be formed by the massless scalar field $Q$ with
the integral of motion $a^2 \! Q'=\sqrt{\rho_0}$.

 In terms of
the standard cosmological definitions of the
redshift\cite{Behnke_02} \bea \label{denspar}
1+z \equiv \frac{1}{a(\eta)}~.
\eea
 the density parameter
 $\Omega_c(z)=\sum\limits_{J=-2,0,1,4}^{}\Omega_{J}a^J$ in Eq. (\ref{rh}) takes the form
 \be \label{coc} \Omega_c(z)=\Omega_{\rm
rig}(1+z)^2+\Omega_{\rm rad}+ \frac{\Omega_{m}}{(1+z)} +
\frac{\Omega_{\Lambda}}{(1+z)^4}~. \ee 

 Then the equation~(\ref{cce}) takes the form
\be \label{etad}
H_0\frac{d\eta}{dz}=\frac{1}{(1+z)^2}\frac{1}{\sqrt{\Omega_c(z)}}~,
\ee and determines the dependence of the conformal time on the
redshift factor. Recall this 
conformal time - redshift relation is valid in both the SC and CC,
where this conformal time is used for description of a light
ray\cite{ps1,pp}.

A light ray traces a null geodesic, i.e. a path for which the
conformal interval $(ds^L)^2=0$ thus satisfying the equation
${dr}/{d\eta} = 1$. As a result we obtain for the coordinate
distance $r$ as a function of the redshift \be \label{rdi} H_0
r(z)=\int_0^z \frac{dz'}{(1+z')^2}\frac{1}{\sqrt{\Omega_c(z')}}.
\ee
 This coordinate
distance -- redshift relation (\ref{rdi}) is a basis of the
luminosity distance -- redshift relation in SC.
 The derivation of luminosity-distance -- redshift relation in CC
 is based on
 the calculation of this relation in SC and the conformal
 postulate (\ref{c-s}) and (\ref{cc-1ell}).

 In order to calculate  the SC luminosity-distance -- redshift relation
 consider a distant source of photons. In conformal coordinates,
  photons behave exactly
as in Minkowski space. Hence, conformal times between emissions of
two subsequent photons and absorptions of these photons are equal.
(This is true both in SC and CC.) The number of photons emitted
per unit conformal time and absorbed in unit conformal time by a
detector covering entire sphere around the source is the same
irrespectively of the position of this sphere,
 \be\label{r-1}
 \frac{dn}{d\eta_{\rm abs}}=\frac{dn}{d\eta_{\rm emis}}
 \ee
Physical times in both cosmologies between the emission emissions
and absorptions are different, in SC because $dt = ad\eta$, and in
CC because time is measured by clocks of mass const$/a$. Hence
 \be\label{r-2}
 \frac{dn}{d\eta_{\rm abs}}=\frac{a(z)}{a_0}\frac{dn}{d\eta_{\rm emis}}
 \ee
where $a(z)$ is the scale factor at emission, and $a_0=1$ is the
scale factor at absorption.  This formula is valid both in SC and
in CC.

The second effect is redshift. Overall, the energy flow through
entire sphere is
 \be\label{r-3}
 \frac{d\varepsilon}{dt}=\frac{1}{1+z}\frac{1}{1+z}L
 \ee
where the first factor is due to redshift, the second factor is
due to the effect  (\ref{r-2}), and $L$ is absolute luminosity of
the source. This formula is valid both in SC and in CC.

Finally, to obtain visible SC luminosity, one has to divide
(\ref{r-3}) by the present area of the sphere. 
This gives for the visible SC luminosity is equal to  energy of
per unit time and per unit surface, \be\label{r-4}
 \frac{d\varepsilon}{dsdt}=\frac{1}{1+z}\frac{1}{1+z}L\frac{1}{4\pi r(z^2)}
 \ee
Defining, as usual, the SC luminosity-distance $\ell_s$ such that
\be\label{r-5}
 \frac{d\varepsilon}{dsdt}=\frac{1}{4\pi \ell_s^2}L
 \ee
 one finds the SC luminosity-distance -- redshift relation
  \be \label{scr} \ell_{\rm s}(z) =(1+z)^2 r_s
 = (1+z)  r(z)~, \ee
 The CC luminosity-distance -- redshift relation
  can be obtained
 from (\ref{cc-1ell})
    in accord
  with the cosmological
  postulate (\ref{c-s})
 \be \label{ccr} \ell_{\rm c}(z) =(1+z)\ell_{\rm s}(z)  =(1+z)^2  r(z)~. \ee
 because all measurable lengths in CC and SC differ, and
all observable  lengths (\ref{cc-1}) in CC (\ref{C-1}) contain an
additional factor $(1+z)$.

\section{\label{s-2} Magnitude-Redshift Relation}

Typically to test cosmological theories one should check a relation
between an apparent magnitude and a redshift. In both SC and CC
models it should be valid the effective magnitude-redshift relation:
\be \mu (z)\equiv m(z)- M = 5 \log{[H_0\ell(z)]} + {\mathcal M},
\label{magz} \ee
 where
$m(z)$ is an observed magnitude, $M$ is the absolute magnitude,
${\mathcal M}$ is a constant with recent experimental data for
distant SNe. Values of $\mu_i$, $z_i$ and $\sigma_i$ could be taken
from observations of a detected supernova with index $i$
($\sigma_i^2$ is a dispersion for the $\mu_i$ evaluation). Since we
deal with observational data  we should choose model parameters to
satisfy an array of relations (\ref{magz}) by the best way because
usually, a number of relations is much more than a number of model
parameters and there are errors in both theory and observations (as
usual we introduce indices for the relations corresponding to all
objects). Typically, $\chi^2$-criterium is used to solve the
problem, namely, we calculate
\be \chi^2 = \sum_i \frac{(\mu_i^{\rm
theor}-\mu_i)^2}{\sigma_i^2}, \label{chi2} \ee
where $\mu_i^{\rm
theor}$ are calculated for given $z_i$ with the assumed theoretical
model and after that we can evaluate the best fit model parameters
minimizing $\chi^2$-function.


\section{Total sample analysis}

For the standard cosmological model for the 186 SNe (the "gold" and
"silver" sample),\footnote{To express differences in quality of
spectroscopic and photometric data the supernovae were separated
into "high-confidence" ("gold") and "likely but not certain"
("silver") subsets.\cite{Riess_04}}  a minimum of the
$\chi^2$-function gives us $\Omega_m=0.28$ ($\chi_{\rm
SC~flat}^2=232.4$) and $\Omega_m=0.31, \Omega_\Lambda=0.80$ assuming
$|\Omega_k| \leqslant 0.11$ ($\chi_{\rm SC~flat}^2=231.0$). Since
other cosmological tests dictate that the Universe should be almost
flat and $\Omega_m=0.28$ is an acceptable value,\cite{Will_2006} we
choose the flat SC model for a reference.

\begin{figure}[t!]
\begin{center}
\includegraphics[width=10.5cm]{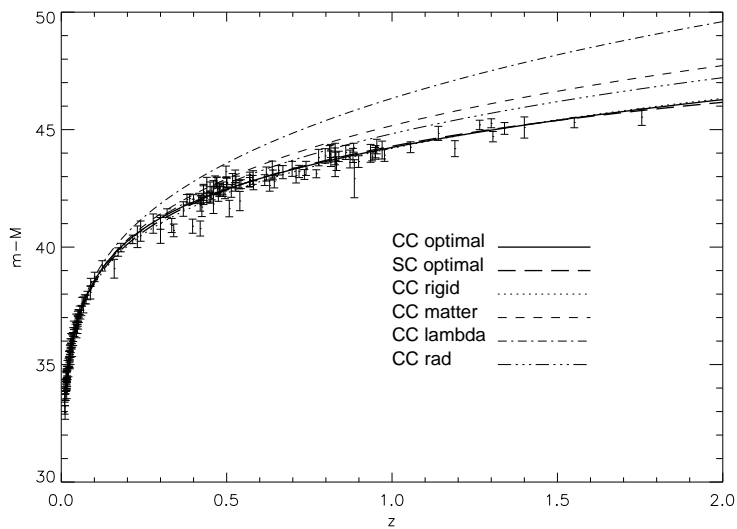}
\end{center}
\caption{$\mu(z)$-dependence for  cosmological models in SC and CC.
The data points include 186 SN Ia (the "gold" and "silver" sample)
used by the cosmological supernova HST team.
For a
reference we use the best fit for the flat standard cosmology model
with $\Omega_m=0.27, \Omega_\Lambda=0.73$ (the thick dashed line),
the best fit for CC is shown with the thick  solid line. For this CC
model we do not put any constraints on $\Omega_m$.} \label{fig1}
\end{figure}

\begin{table}[t!]
\tbl{The fits for CC models for the total sample with different
constraints on $\Omega_m$ (the best fits are shown in first, second
and third rows, two almost best fits are presented in fourth and
fifth rows).}
{
\begin{tabular}{|c|c|c|c|c|c|}
\hline Constraints on $\Omega_m$ & $\Omega_m$ & $\Omega_\Lambda$ & $\Omega_{\rm rad}$ & $\Omega_{\rm rig}$ & $\chi^2$\\
\hline \hline
No constraints             & -4.13      & 3.05             & 0.05               & 2.085              & 226.64 \\
\hline
$\Omega_m \ge 0.$          & 0.         & 0.18             & 0.                 & 0.80               & 242.76 \\
\hline
$0.2 \le \Omega_m \le 0.3$ & 0.2        & 0.013            & 0.                 & 0.75               & 244.67\\
 \hline
$0.2 \le \Omega_m \le 0.3$ & 0.29        & 0.0            & 0.                 & 0.7               & 246.58\\
 \hline
 $0.2 \le \Omega_m \le 0.3$ & 0.27        & 0.0            & 0.                 & 0.72               & 245.66\\
 \hline
\end{tabular}\label{tabl1}}
\end{table}

\begin{table}[t!]
\tbl{The $\chi^2$ values for pure flat CC models for the total
sample. The models are shown in Figs. \ref{fig1},\ref{fig2} as
references.}
{\begin{tabular}{|c|c|c|c|c|c|}
\hline
Model types  &$\Omega_m=1$ & $\Omega_\Lambda=1$ & $\Omega_{\rm rad}=1$ & $\Omega_{\rm rig}=1$ \\
\hline \hline
$\chi^2$        & 924.27      & 4087.93             & 478.42               & 276.71      \\
\hline
 \end{tabular}
\label{tabl2}
}
\end{table}

In Fig.~\ref{fig1} we compare the SC and CC fits  for the effective
magnitude-redshift relation if we will not put any constraint on
$\Omega_m$ (in this case we assume that SNe Ia data is the only
cosmological test for CC models we obtain the best fit expressed in
the first row in Table~\ref{tabl1}). Analyzing the curves
corresponding to the best fits for SC and CC models one can say that
they almost non-distinguishable, moreover the best fit CC provide
even better the $\chi^2$ value (see first row in Table \ref{tabl1}).
We would not claim that we discovered a cosmological model with
negative $\Omega_m$, but we would like to note that the best CC and
SC fits are almost non-distinguishable from a formal statistical
point of view (the thick solid and long dashed lines, respectively
in Fig.~\ref{fig1}).
An appearance of the fit with negative $\Omega_m$ can be caused also by
systematical errors in observational data.
Sometimes new physical phenomena are introduced
qualitatively with  the same statistical arguments (such as an
introduction of the phantom energy, for instance), but if we should
follow a more conservative approach, we could conclude that in this
case we should simply put extra constraints on $\Omega_m$ to have no
contradictions to other cosmological (and astronomical) tests.
 So,
if we put "natural" constraints on $\Omega_m \geqslant 0$, the best
fit parameters for CC model are presented in second row in Table
\ref{tabl1}. In this case the $\chi^2$ difference between two CC
models ($\Delta \chi^2 \thickapprox 16$) is not very high and a
difference between this fit and the SC best fit for a flat model is
about $\Delta \chi^2 \thickapprox 10$ (or less than 5\%), it means
the CC fit is at an acceptable level. For references, we plotted
also pure flat CC models, so that rigid, matter, lambda and
radiation models are shown with thin dotted, short dashed, dot dash,
dash dot dot dot lines, respectively. Corresponding $\chi^2$ values
are given in Table~2. One can see that only pure flat rigid CC model
has relatively low $\chi^2$ values (and it could be accepted as a
rough and relatively good fit for cosmological SNe Ia data), but
other models should be definitely ruled out by the observational
data.

\begin{figure}[t!]
\begin{center}
\includegraphics[width=10.5cm]{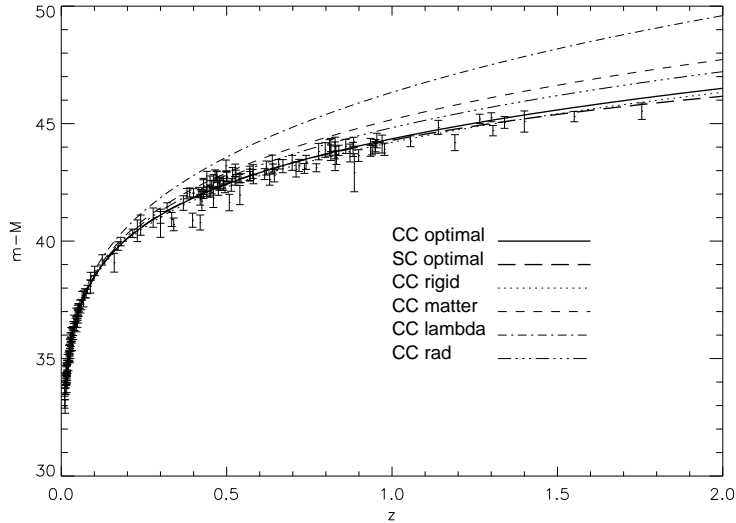}
\end{center}
\caption{$\mu(z)$-dependence  cosmological models in SC and CC. As
in previous figure, the data points include 186 SN Ia (the "gold"
and "silver" sample) used by the cosmological supernova HST team
and for a reference we use the best fit for the flat
standard cosmology model with $\Omega_m=0.27, \Omega_\Lambda=0.73$
(the thick dashed line), the best fit for CC is shown with the thick
solid line. For this CC model we assume $\Omega_m \in [0.2,0.3]$.}
\label{fig2}
\end{figure}

So, if we put further constraints on $0.2 \leqslant \Omega_m
\leqslant 0.3$ based on measurements of clusters of galaxies and
other cosmological arguments,\cite{Bahcall_98} the best fit
parameters for CC model are presented in third row in Table~1. In
this case the $\chi^2$ difference between two CC models ($\Delta
\chi^2 \thickapprox 18$) is not very high also and a difference
between $\chi^2$ for the CC  and  SC models is about $\Delta \chi^2
\thickapprox 12$ (or about 5\%), it means the CC fit is at an
acceptable level. Dependence of $\chi^2$ on $\Omega_m$ is very weak
and we present intermediate fits for CC model in fourth and fifth
rows in Table~1 (there is a valley of $\chi^2$ function in the
$\Omega_m$ direction). The best fit for a CC model with parameters
given in third row in Table~1 is shown as the optimal fit for the CC
model in Fig.~\ref{fig2} with the solid thick line. Other lines are
the same as in Fig.~\ref{fig1} and they are shown for references.

\section{Analysis for more recent "gold" SNe Ia  data}

In the section we describe  results of our analysis for more recent
"gold" data (for 192 SNe Ia)\cite{Davis_07} (where the authors collected observational data published earlier
\cite{Wood_07,Riess_07}). These SNe Ia are shown in Fig. \ref{fig3}.
For the standard flat cosmological model $\chi^2 = 196.1$, meanwhile for
the best fit for CC model $\chi^2 = 203.03$, so the difference for these two approximations
is only few percent, therefore

\begin{figure}[t!]
\begin{center}
\includegraphics[width=10.5cm]{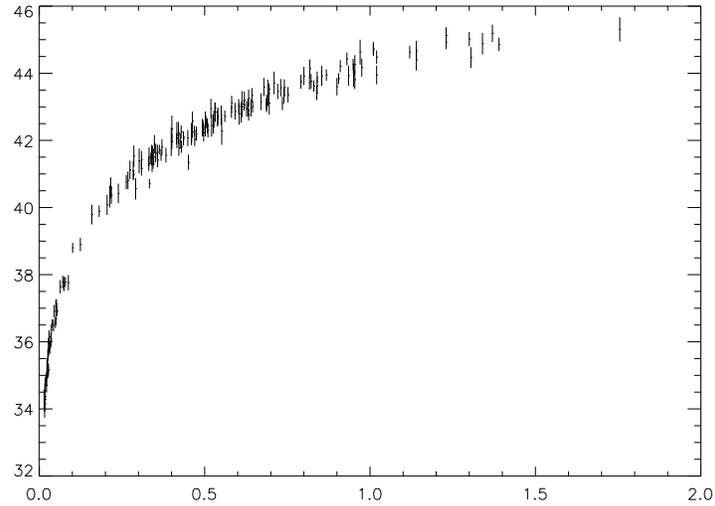}
\end{center}
\caption{Observational data without theoretical fits.}
\label{fig3}
\end{figure}

\begin{figure}[t!]
\begin{center}
\includegraphics[width=10.5cm]{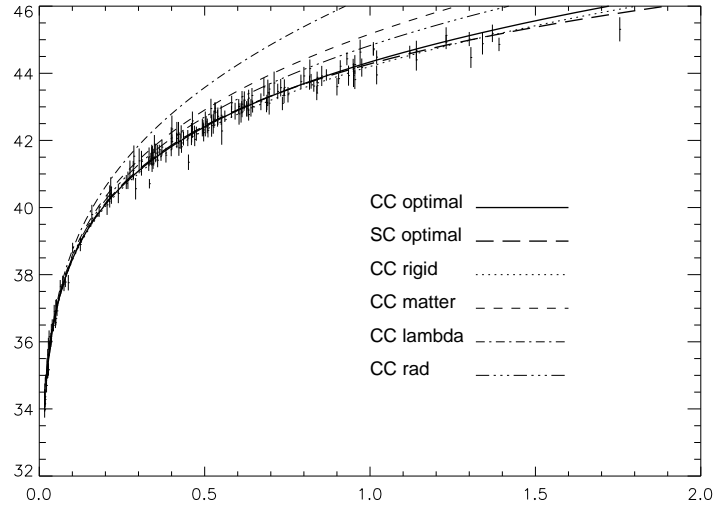}
\end{center}
\caption{$\mu(z)$-dependence  cosmological models in SC and CC for Davis et al. data.}
\label{fig4}
\end{figure}

\begin{table}[t!]
\tbl{The fit for CC models for the total Davis et al. sample  without
constraints on $\Omega_m$.}
{
\begin{tabular}{|c|c|c|c|c|c|}
\hline Constraints on $\Omega_m$ & $\Omega_m$ & $\Omega_\Lambda$ & $\Omega_{\rm rad}$ & $\Omega_{\rm rig}$ & $\chi^2$\\
\hline \hline
No constraints             & .20      & .03             & 0.00              & 0.81              & 203.03 \\
\hline
\end{tabular}
\label{tabl1a}}
\end{table}

\begin{table}[t!]
\tbl{The $\chi^2$ values for pure flat CC models for the total
sample.}
{\begin{tabular}{|c|c|c|c|c|c|}
\hline
Model types  &$\Omega_m=1$ & $\Omega_\Lambda=1$ & $\Omega_{\rm rad}=1$ & $\Omega_{\rm rig}=1$ \\
\hline \hline
$\chi^2$        &1312.74
      & 6350.61              &  590.60               & 238.62      \\
\hline
 \end{tabular}
\label{tabl2a}
}
\end{table}

\section{\label{s-3}Conclusions}

Using "gold" and "silver" 186 SNe Ia\cite{Riess_04} we confirm in
general and clarify previous conclusions about CC model parameters,
done earlier with analysis of smaller sample of SNe Ia data\cite{Behnke_02,Behnke_04}
 that the pure flat rigid CC model could
fit the data relatively well since $\Delta \chi^2 \thickapprox 44.3$
(or less than 20~\%) in respect of the standard cosmology flat model
with $\Omega_m=0.28$. Other pure flat CC models should be ruled out
since their $\chi^2$ values are too high.

For the total sample, if we consider CC models with a "realistic"
constraint $0.2 \leqslant \Omega_m \leqslant 0.3$ based on other
astronomical or cosmological arguments except SNe Ia data, we
conclude that the standard cosmology flat model with $\Omega_m=0.28$
is still preferable in respect to the  fits for the CC models (with
$\Omega_m=0.2$ and $\Omega_{\rm rig}=0.75$ or $\Omega_m=0.27$ and
$\Omega_{\rm rig}=0.72$, for instance, see third and fifth rows in
Table~1),  but the preference is not very high (about 5~\% in
relative units of $\chi^2$ value), so the CC models could be adopted
as acceptable ones taking into account possible sources of errors in
the sample and systematics.

Thus, for  CC model fits calculated with SNe Ia data, in some sense,
a rigid equation of state could substitute the $\Lambda$-term (or
quintessence) in the Universe content. As it was mentioned above the
rigid matter can be formed by a free massless scalar field.

The best CC models provide almost the same quality fits of SNe Ia
data as the best fit  for the SC flat model, however the last
(generally accepted) model is more preferable.


\section*{Acknowledgements}

The authors are indebted to B.M.~Barbashov, J.~Wang, J.~Zhang,
G.~Zhou for important discussions.

 AFZ  is grateful  to the National Natural Science Foundation
of China (NNSFC)  (Grants \# 10703007, 10873020, G10573025,
40674081, 10603008, 40890161, and 10733020) and National Basic
Research Program of China (G2006CB806303) for a partial financial
support of the work.


\begin{thebibliography}{999}

\bibitem{Riess_04}
A.D. Riess, L.-G. Strolger, J.~Tonry et al., {\it Astrophys. J.}, {\bf 607}, 665
(2004).
\bibitem
{Will_2006}
C.~Will, {\it Living Rev. Relativity,} {\bf  9}, 3 (2006);\\
S.~Bludman, astro-ph/0605198; \\
E.J.~Copeland, M.~Sami, S.~Tsujikawa, {\it Int. J. Mod. Phys. D}, {\bf 15}, 1753 (2006); hep-th/0603057; \\
N.~Straumann, {\it Mod. Phys. Lett. A}, {\bf 21}, 1083 (2006).

\bibitem{Zakharov_06}
A.F. Zakharov, A.A. Nucita, F. De Paolis, G. Ingrosso,
{\it Phys. Rev. D} {\bf 74},  107101 (2006);\\
  X. Jin, D. Liu, X. Li,
 astro-ph/0610854.

\bibitem{Panagia_05}
N.~Panagia,  astro-ph/0502247.
\bibitem{Behnke_02}
D.~Behnke,  D.B.~Blaschke, V.N.~Pervushin, D.V.~Proskurin, {\it
Phys. Lett. B}  {\bf 530}, 20 (2002).

\bibitem{Behnke_04}
D.~Behnke,  {\it Conformal Cosmology Approach to the Problem of
Dark Matter,} PhD Thesis, Rostock Report MPG-VT-UR 248/04 (2004).


\bibitem{CC_papers}
D.B.~Blaschke, S.I.~Vinitsky, A.A.~Gusev, V.N.~Pervushin, and
D.V.~Proskurin, {\it Phys. Atom. Nucl.} {\bf  67}, (2004)
 1050;\\
B.M.~Barbashov, V.N.~Pervushin, A.F.~Zakharov, V.A.~Zinchuk,
{\it Int. J. Mod. Phys. A}, {\bf 12}, 5957 (2006);
astro-ph/0511824;\\
 B.M.~Barbashov, V.N.~Pervushin, A.F.~Zakharov,
V.A.~Zinchuk,
{\it Int. J. Geom. Meth.
Mod. Phys.}  {\bf 4}, 171 (2007); hep-th/0606054;\\
 B.M.~Barbashov, V.N.~Pervushin, A.F.~Zakharov, V.A.~Zinchuk,
{\it Phys. Atom. Nucl.} {\bf 70} 191 (2007); \\
A.F.~Zakharov, V.N.~Pervushin, V.A.~Zinchuk,
{\it Phys. Part. and Nucl.} {\bf  37}, 104 (2006);\\
B.M.~ Barbashov, V.N.~Pervushin, A.F.~Zakharov, V.A.~Zinchuk,
 {\it Physics Letters B}  {\bf 633}, 438  (2006); \\
V.N.~Pervushin, A.F.~Zakharov, V.A.~Zinchuk,
Cosmic Evolution as
"Superfluid" Motion in General Relativity,
in {\it Proc. the INTAS Summer School and International
Conference "New Trends in  High-Energy Physics (experiment,
phenomenology, theory)"}, Yalta, Crimea, Ukraine, 2005, Bogoliubov
Institute for Theoretical Physics, National Academy of Sciences of
Ukraine, Joint Institute for Nuclear Research (Dubna), p.~271;\\
 B.M.~Barbashov, V.N.~Pervushin, A.F.~Zakharov, V.A.~Zinchuk,
 Hamiltonian
General Relativity in Finite Space and Cosmological Potential
Perturbations,
in {\it Proc. "Nuclear Science and Safety in Europe"}, T. Cechak et
al. (eds.), Springer, 2006, p. 125; astro-ph/0511824; \\
V.N.~Pervushin, A.F.~Zakharov, V.A.~Zinchuk,
Cosmic Evolution as "Superfluid" Motion in General Relativity,
in {\it Proc. "Nuclear Science and Safety in Europe"}, T. Cechak et
al. (eds.), Springer, 2006, p. 201;  \\
B.M.~Barbashov, V.N.~Pervushin, A.F.~Zakharov, V.A.~Zinchuk,
Einstein - Hilbert Formulations of GR and Modern Cosmology,
in {\it Proc.  of the 8th International Workshop Relativistic Nuclear
Physics: from Hundreds MeV to TeV}, (JINR, Dubna, Russia, 2006) p.~11;\\
B.M.~Barbashov, V.N.~Pervushin, A.F.~Zakharov, V.A.~Zinchuk,
Quantum Gravity as Theory of ``Superfluidity'', in  {\it Proc. of
the XXVIII Spanish Relativity Meeting E.R.E. "A Century of
Relativity Physics"} Oviedo (Asturias) Spain, (American Institute
of Physics, 2006), {\bf 841}, p.~362.

\bibitem{lich}
A.~Lichnerowicz, {\it Journ. Math. Pures and Appl.}, {\bf B 37}, 23
(1944); \\
J.W.~York. (Jr.), {\it Phys. Rev. Lett.}, {\bf 26}, 1658 (1971);
 K.~Kuchar, {\it J. Math. Phys.}, {\bf 13}, 768 (1972).


\bibitem{Riess_2001}
 A.G.~Riess {\it et al.}, {\it Astrophys. J.} {\bf 560}, 49 (2001);\\
D.~Puetzfeld, {\it Class. Quant. Grav.}, {\bf 19}, 4463 (2002);\\
P.P.~Avelino, C.J.A.P.~Martins, {\it Astrophys. J.} {\bf 565}, (2002);\\
N.~Dalal, K.~Abazajian, E.~Jenkins, A.V.~Manohar, {\it
Phys.Rev.Lett.}
{\bf 87}, 141302 (2001);\\
M.~Tegmark, {\it  Phys. Rev.D}, {\bf 66} 103507 (2002);\\
Z.-H.~Zhu, M.~Fujimoto, {\it Astrophys. J.}, {\bf 585}, 52 (2003).


\bibitem{Dirac_73}
P.A.M. Dirac, {\it Proc. R. Soc. Lond.}, {\bf A 333},  403 (1973).


\bibitem{weyl} H.Weyl, {\it Sitzungsber. d. Berl. Akad.} 465 (1918).
\bibitem{Faraoni_99}
V. Faraoni, {\it Phys. Rev. D}, {\bf 59}, 084021 (1999).

\bibitem{Zee_79}
A. Zee, Phys. Rev. Lett. {\bf 42},  417 (1979);
L. Smolin, Nucl. Phys. {\bf B 160}, 253 (1979);
D.S. Salopek, J.R. Bond and J.M. Bardeen, Phys. Rev. D {\bf 40},  1753 (1989);
R. Fakir and W.G. Unruh, Phys. Rev. D {\bf 41} 1792 (1990).




\bibitem{Riess_98}
A.G.~Riess  et al., {\it Astron. J.} {\bf  116}, 1009 (1998);\\
S.~Perlmutter  et al., {\it Astrophys. J.} {\bf  517}, 565 (1999).




\bibitem{Quintessence}
R.R.~Caldwell, R.~Dave, P.J.~Steinhardt, {\it Phys. Rev. Lett.} {\bf 80}, 1586 (1998);\\
R.R.~Caldwell,  {\it Phys.  Lett. B} {\bf 545}, 23 (2002);\\
T.~Padmanabhan, astro-ph/0411044; \\
Y.-F. Cai, H.~Li, Y.-S.~Piao, X.-M.~Zhang,
{\it Phys.  Lett. B} {\bf 646}, 141 (2007).


\bibitem{Linde_07}
A.D.~Linde, arxiv.org:0705.0164v2[hep-th]
%

%
%

\bibitem{vp}
V. Pervushin et al., {\it Phys. Lett. B} {\bf 365}, 35 (1996).

\bibitem{ps1}
V.N.~Pervushin, V.I.~Smirichinski, {\it J. Phys. A: Math. Gen.} {\bf 32}, 6191 (1999).

\bibitem{pp}
M.~ Pawlowski, V.N.~Pervushin, {\it Int. J. Mod. Phys. A} {\bf 16}, 1715 (2001).

\bibitem{Bahcall_98}
N.A.~Bahcall, X.~Fan, {\it Publ. Nat. Academy of Science}, {\bf 95}, 5956 (1998);\\
N.A.~Bahcall, X.~Fan, {\it Astrophys. J.} {\bf 504}, 1 (1998);\\
D.N.~Spergel, R.~Bean, O. Dore' et al. astro-ph/0603449;\\
U.~Seljak, A.~Slosar, P.~McDonald, {\it J. Cosm. Astroparticle Phys.} {\bf 10}, 14
(2006).
\bibitem{Davis_07}
T. M. Davis,  E. Mörtsell, J. Sollerman, et al.  {\it Astrophys. J.} {\bf 666} 716 (2007); (astro-ph/0701510).

\bibitem{Wood_07}
W.M. Wood-Vasey, G.  Miknaitis, C. W. {\it Astrophys. J.} {\bf 666} 694 (2007); (astro-ph/0701041).

\bibitem{Riess_07}
A. Riess, L.-G. Strolger, S. Casertano et al. {\it Astrophys. J.} {\bf 659} 98 (2007); (astro-ph/0611572)

\end{thebibliography}
\end{document}